\documentclass[runningheads]{llncs}
\usepackage{graphicx}
\usepackage{subfig}
\usepackage{amsmath}
\usepackage{multirow}
\begin{document}

\title{Cascaded Robust Learning at Imperfect Labels for Chest X-ray Segmentation}
%
\titlerunning{Cascaded Robust Learning at Imperfect Labels}
%
\author{Cheng Xue\inst{} \and
Qiao Deng\inst{} \and
Qi Dou\inst{} \and
Pheng-Ann Heng\inst{}}
\authorrunning{C. Xue et al.}
%
\institute{Chinese University of Hong Kong}
%
\maketitle              
\begin{abstract}
The superior performance of CNN on medical image analysis heavily depends on the annotation quality, such as the number of labeled image, the source of image, and the expert experience. The annotation requires great expertise and labour. To deal with the high inter-rater variability, the study of imperfect label has great significance in medical image segmentation tasks. In this paper, we present a novel cascaded robust learning framework for chest X-ray segmentation with imperfect annotation. 
Our model consists of three independent network, which can effectively learn useful information from the peer networks. The framework includes two stages. In the first stage, we select the clean annotated samples via a model committee setting, the networks are trained by minimizing a segmentation loss using the selected clean samples. In the second stage, we design a joint optimization framework with label correction to gradually correct the wrong annotation and improve the network performance. 
We conduct experiments on the public chest X-ray image datasets collected by Shenzhen Hospital. The results show that our methods could achieve a significant improvement on the accuracy in segmentation tasks compared to the previous methods.

\end{abstract}
\section{Introduction}
Deep neural networks (DNNs) have achieved human-level performance on many medical image analysis tasks, such as melanoma diagnosis \cite{esteva2017dermatologist}, pulmonary nodules detection \cite{setio2017validation}, retinal disease \cite{de2018clinically}, and lumpy node metastases detection \cite{bejnordi2017diagnostic}. These outstanding performances heavily rely on massive training data with high-quality annotations. Annotation of medical images, especially for pixel-level annotation for segmentation tasks, is costly and time-consuming. The process is experience-prone, while the annotations from different clinical experts may have disagreements that are usually inevitable for the blurred boundary of lesions and organs. 

Previous studies show that the DNNs trained by noisy labeled datasets can cause performance degradation. That is because the huge memory capacity and strong learning ability of DNNs can remember the noisy labels and easily overfit to them \cite{zhang2016understanding,zhu2019pick,shu2019lvc}. Hence, it is important to develop DNNs with strong robustness to noisy labels. Many studies have addressed the issue of noisy label in medical analysis community. Goldberger et al.\cite{goldberger2016training} added an additional softmax layer to estimate the correct labels. Xue et al. \cite{xue2019robust} proposed to consider the noisy sample and hard sample by an on-line sample selection module and re-weighting module. Zhu et al. \cite{zhu2019pick} proposed the automatic quality evaluation module and overfitting control module to update the network parameters. Shu et al.\cite{shu2019lvc} presented a LVC-Net losses function by combining noisy labels with image local visual cues  to generate better semantic segmentation. Le et al. \cite{le2019pancreatic} utilized a small set of clean training samples to assign weights to training samples.

To tackle the challenging problem of noisy labeled segmentation masks, we present a cascaded learning framework for lung segmentation using the X-ray images with imperfectly annotated ground truth. In the first stage, our framework selects clean annotated samples according to the prediction confidence and uncertainty of samples, that is inspired by the ideas of Co-teaching \cite{han2018co}. Specifically, our model consists of three independent networks being trained simultaneously, each network is real-time updated according to the prediction results of the other two networks. For a clean annotated sample, the three networks tend to produce high confidence prediction with smaller inter-rater variance. Thus, the samples with close prediction and high confidence are selected as the high-quality sample, which will be only used to contribute in the weight backpropagation process. Since the selection stage leads to a low utilization efficiency of the valuable training data, we propose a label correction module in the second stage, which can correct the imperfect label. Furthermore, a joint optimization scheme is designed to cooperatively supervise the three networks with the original label and the corrected one. Our method was extensively evaluated
on the dataset Shenzhen chest x-ray \cite{jaeger2013automatic,candemir2013lung,stirenko2018chest}. The results demonstrate a good capability of our method to the issue of noisy label, that the cascaded robust learning framework can more accurately perform the lung segmentation comparing to other methods.

\section{Method}
\begin{figure}[t]
    \centering
    \includegraphics[width=0.9\textwidth]{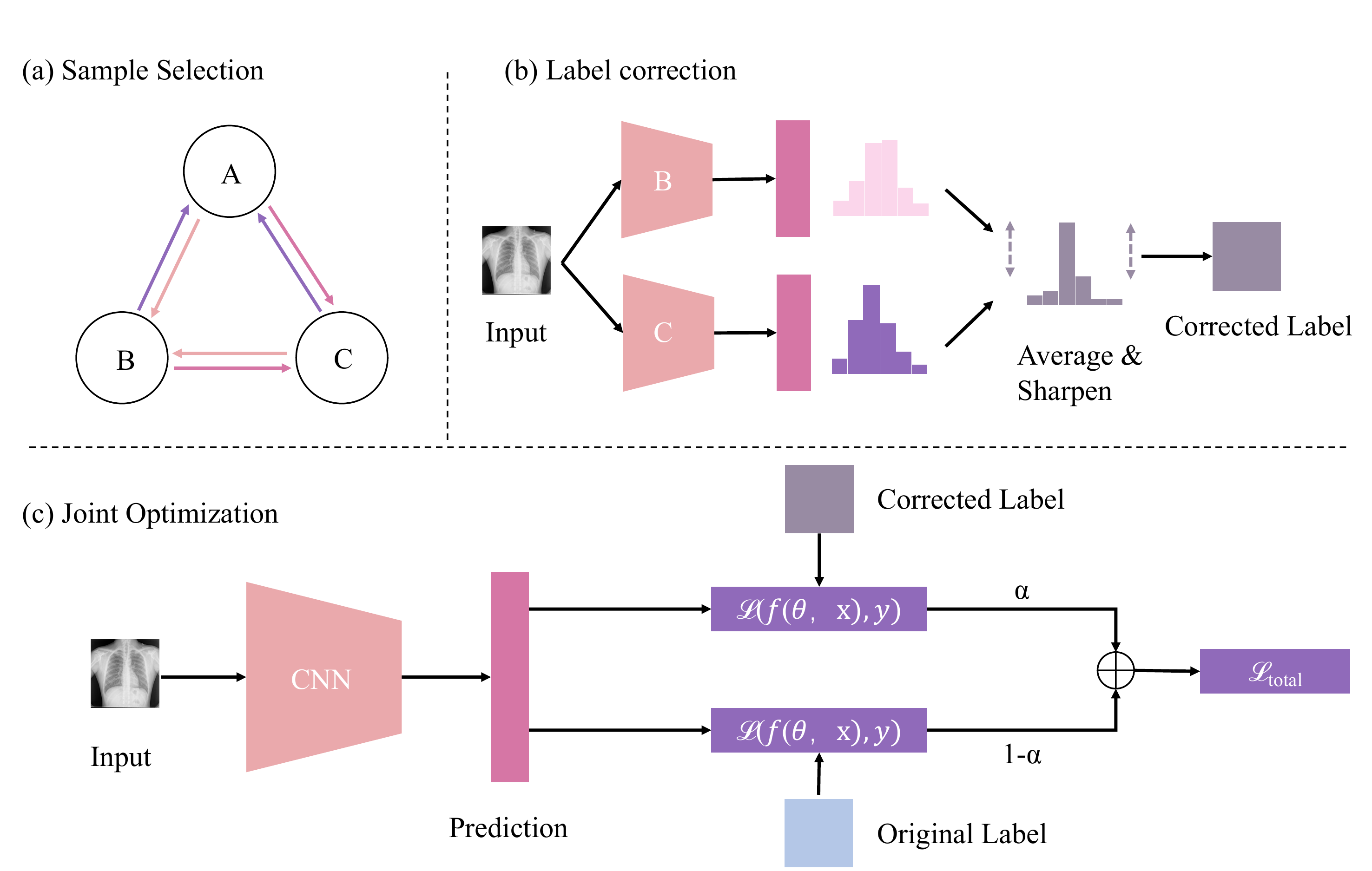}
    \caption{Illustration of the pipeline of our cascaded robust learning framework. (a) shows the first sample selection stage, where three networks trained independently, but updated according to the prediction of the other two peer networks. (b) and (c) are the second stage. (b) shows our proposed label correction module, using the average prediction of two peer networks followed by a sharpening function to produce corrected label $\bar{y}$. (c) shows the joint optimization scheme, the network is supervised by the original label $\hat{y}$ and the corrected label $\bar{y}$.  The final output is given by the average of the three networks.}
    \label{fig:my_label}
\end{figure}
Fig 1. illustrates the framework of our cascaded robust learning method. In the first stage the sample selection module filters the clean samples and update the three networks with the selected clean samples. In the second stage, our method start to correct the imperfect labels, then use the corrected label and original label to jointly optimize the three networks.

\subsection{Sample Selection Stage}
We study the task of chest x-ray segmentation, where the training set contains images $x$ and noisy labeled ground truth $\hat{y}$, while the clean ground truth $y$ is unknown. The goal for this fully supervised segmentation task is to minimize the following object function:
\begin{equation}
    \min_{\theta} \sum_{i=1}^N \mathcal{L}(f(x_i;\theta),\hat{y}_i)
\end{equation}
where $\mathcal{L}$ denotes the loss function (e.g., cross-entropy loss) to evaluate the quality of the network output on inputs. $f(\theta)$ denotes the segmentation neural network with weights $\theta$.

Recent studies show that by updating the network with high confidence samples can improve the robustness to noisy labels \cite{ren2018learning,jiang2017mentornet,han2018co}. Therefore, we propose a novel sample selection framework (SS) to select high confidence samples as the useful training instances. Our framework consisted by three independent networks, where they have identical architecture. We adopt the vanilla U-Net\cite{ronneberger2015u} as the classifier in our experiment. In the training process, we select the high confidence samples with small uncertainty to update each network, because those samples are more likely to be clean labeled instances. In our experiment, we select half batch data as useful information. Concretely, the three networks feed forward and predict the same mini-batch of data. Then for each network, the useful samples for weight updating is obtained by the other two networks as shown in Fig. \ref{fig:my_label}(a). Taking network A as an example, the useful sample for network A is obtained from network B and C, where we first filter out the high uncertainty samples by exclude the ones showing disagreed prediction, then among the low uncertainty samples, the small loss samples was further selected as useful samples for network A. We calculate the uncertainty according to equation \ref{Eq1}.
\begin{equation}
\label{Eq1}
    \mu=|\mathcal{L}(f_B(x_i;\theta_B),\hat{y}_i) - \mathcal{L}(f_C(x_i;\theta_C),\hat{y}_i)|
\end{equation}
where $\mathcal{L}$ denotes the cross-entropy loss. $f_2$ and $f_3$ denote the network B and network C. $\theta_B$ and $\theta_C$ represent the weight of network B and C.

\subsection{Joint Optimization with Label Correction}
In the stage of sample selection, only partial samples can be used for training, where it does not take fully advantage of the imperfect training data. Therefore, we design a joint optimization (JO) framework to train the network with the original label and corrected label, so that the utilization efficiency of training data can be maintained. In order to correct the noisy label, we design an label correction module to work together with the joint optimization scheme.

\paragraph{Label correction}
The sample selection stage first trains an initial network by using image x with noisy label $\hat{y}$. Then we proceed to the label correction phase, as shown in Figure \ref{fig:my_label} (b). We compute the average of three model’s prediction in each iteration, that is followed by an entropy minimization step widely adopted in semi-supervised learning \cite{berthelot2019mixmatch}. Specifically, for the average prediction of the three models, we apply a sharpening function to reduce the entropy of the per pixel label distribution through adjusting the temperature. The sharpening function is the equation \ref{eq2}.
\begin{align}
\label{eq2}
\begin{split}
 q=\frac{1}{2}( f_B((x_i;\theta_B),\hat{y}_i) + f_C((x_i;\theta_C),\hat{y}_i))\\
sharpen(q,T)_i =q_i^{\frac{1}{T}}/\sum_{j=1}^{L}q_j^{\frac{1}{T}}
\end{split}
\end{align}
where $q$ is the average prediction feature map over two models, $T$ is a hyperparameter that adjusts the temperature. As $T$ closes to zero, the output of $Sharpen(q, T)$ will approach a one-hot distribution. Since we will use $q =Sharpen(q, T)$ as a corrected target for the model’s prediction later, $T=0.5$ is chosen to encourage the model to produce lower-entropy prediction.

\paragraph{Joint optimization}
We start the joint optimization stage after $k$ epochs of sample selection. For each uncertain sample, we produce a corrected label for the imperfect input by the label correction module. The corrected label is used in the training process together with the original label as a complementary supervision to jointly supervise the network, as shown in  Equation \ref{eq:joint}. 

\begin{equation}
    \mathcal{L}_{total}=\alpha \times \mathcal{L}(f(x_i;\theta),\hat{y}_i) +(1-\alpha) \times \mathcal{L}(f(x_i;\theta),\bar{y}_i)
\label{eq:joint}
\end{equation}
where $\mathcal{L}$ is the cross entropy loss, $\hat{y}$ is the original noisy label, and $\bar{y}$ is the corrected label produced by the label correction phase. The weight factor $\alpha$ controls the important weight of the two terms, we set $\alpha=0.5$ in our study.

\section{Experiments}
\subsection{Dataset and Pre-processing. }
We evaluated our method on the public Shenzhen chest x-ray datasets \cite{jaeger2013automatic,candemir2013lung,stirenko2018chest}, the segmentation mask were prepared manually by Computer Engineering Department, Faculty of Informatics and Computer Engineering, National Technical University of Ukraine. The dataset contains 566 chest x-ray images and each image has the left and the right lungs. We split the 566 chest x-ray images into 396 images for training and 170 for evaluation. All the images were resized to $256 \times 256$, and normalized as zero mean and unit variance.
\subsection{ Implementation}
The framework was implemented in PyTorch, using a TITAN Xp GPU. We used the SGD optimizer to update the network parameters with weight decay of 0.001 and momentum of 0.9. We adopt an exponential learning rate with initial learning rate set as 0.001. We totally trained 100 epochs, the batch size was 32. We adopted the data augmentation including randomly rotation and randomly horizontal flipping. In order to produce noisy label for the training data, we randomly selected 25$\%$, 50$\%$, 75$\%$ samples from the training set to erode or dilate with the number of iterations $n$ between $5-15$ ($5 \leq n \leq 15$). We adopted the dice coefficient as evaluation criteria for segmentation accuracy evaluation. Fig.2 shows the example of some noisy annotation of the segmentation mask.
\begin{figure}[t]
\begin{tabular}{cccc}
  \includegraphics[width=0.24 \textwidth]{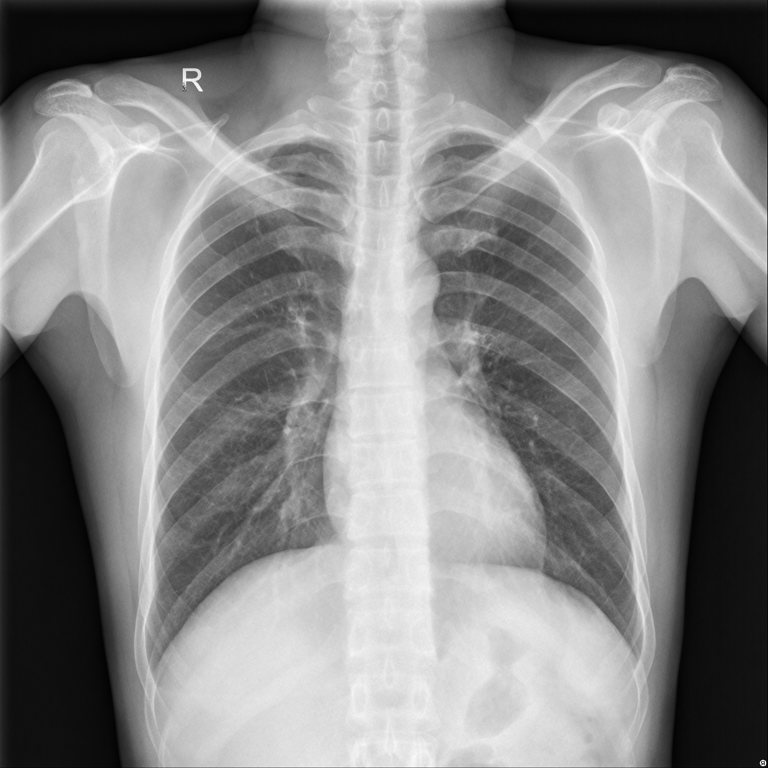} &   \includegraphics[width=0.24 \textwidth]{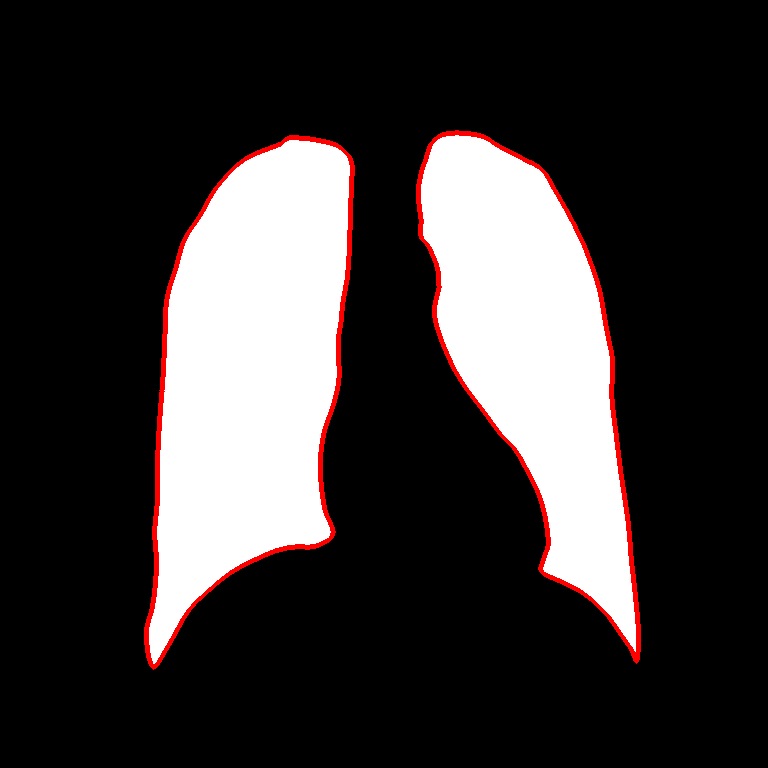}&
  \includegraphics[width=0.24 \textwidth]{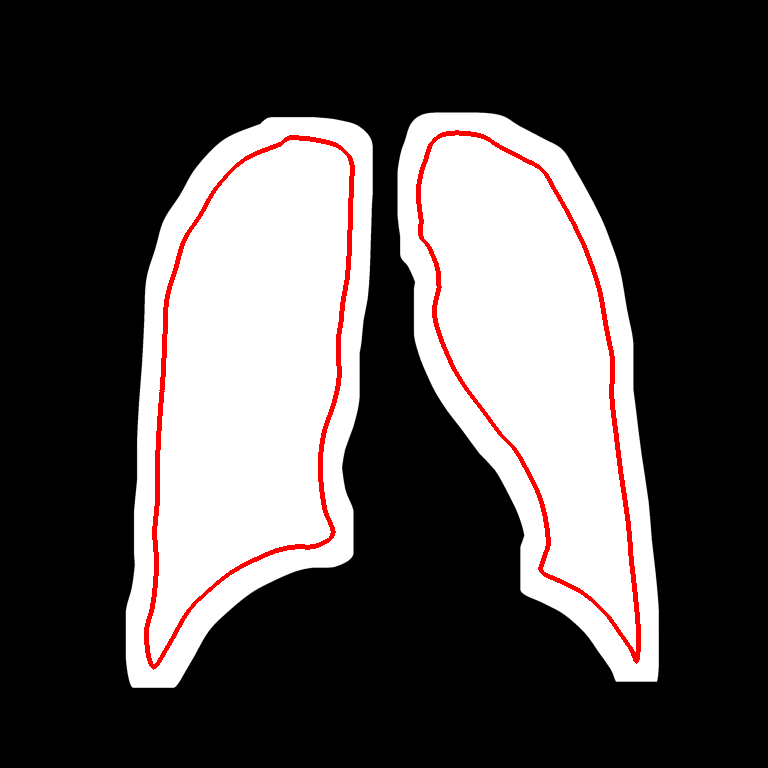}&
  \includegraphics[width=0.24 \textwidth]{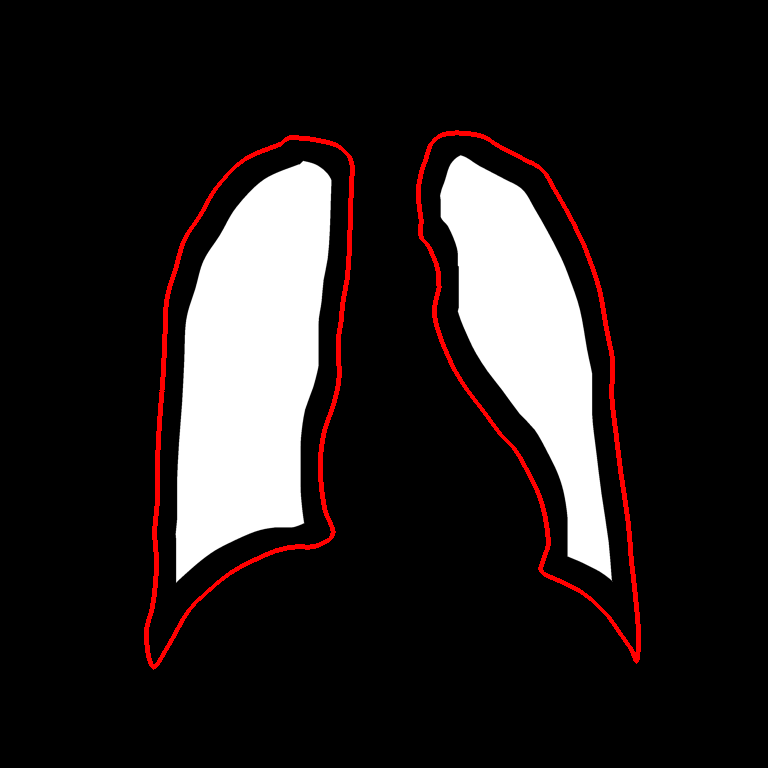}\\
(a) chest X-ray & (b) ground truth &(c)dilation  & (d)erosion\\
\end{tabular}
\caption{Examples of the noise annotation }
\end{figure}

\begin{figure}[t]
\begin{tabular}{cccc}
  \includegraphics[width=30mm]{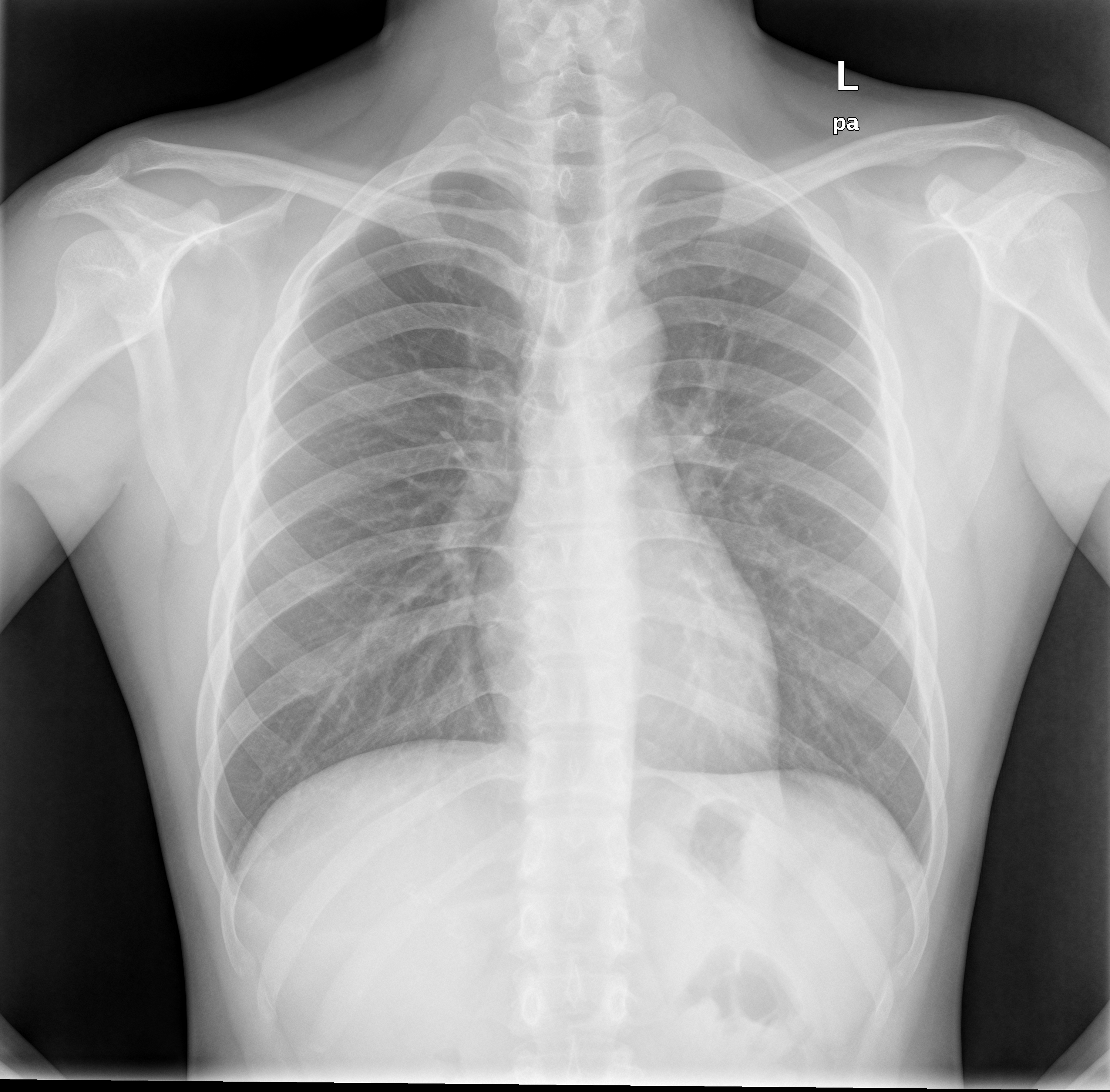} &   \includegraphics[width=30mm]{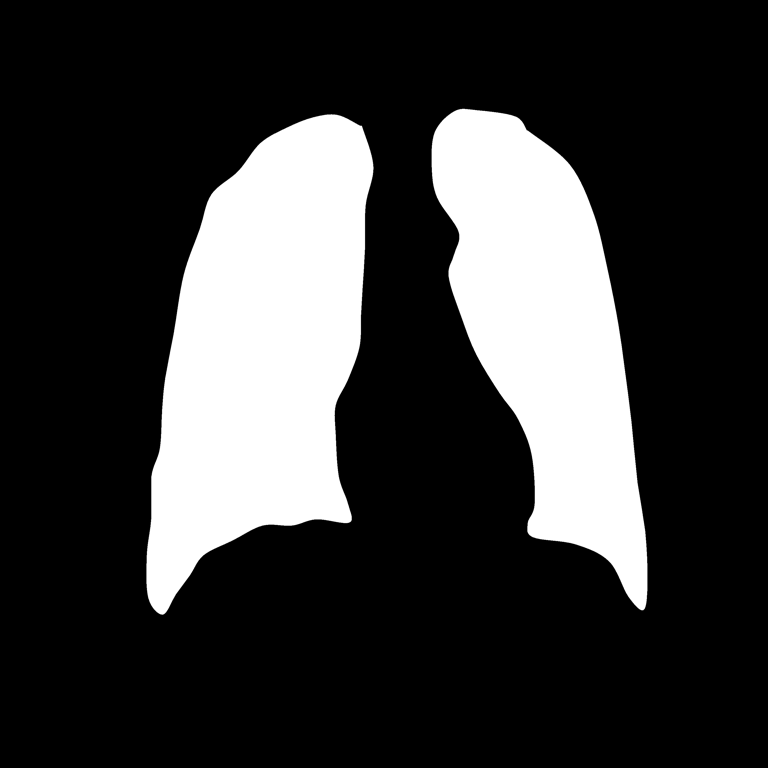}&
  \includegraphics[width=30mm]{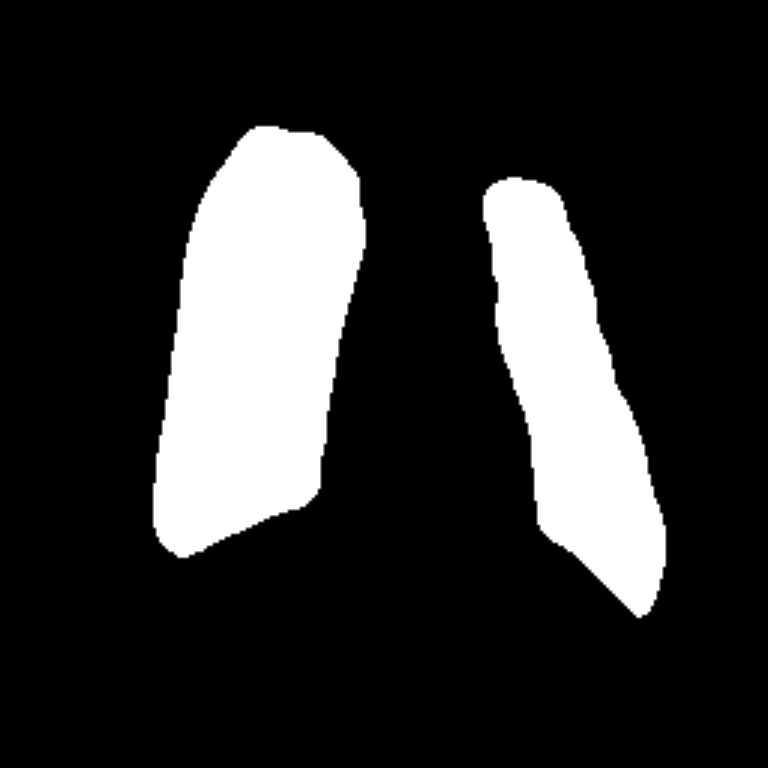}&
  \includegraphics[width=30mm]{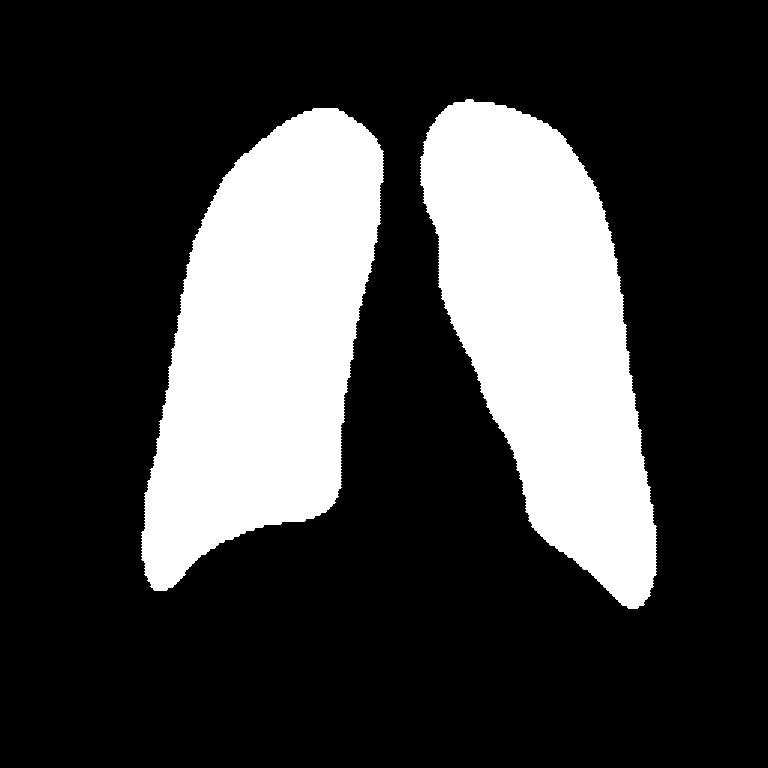}\\
  \includegraphics[width=30mm]{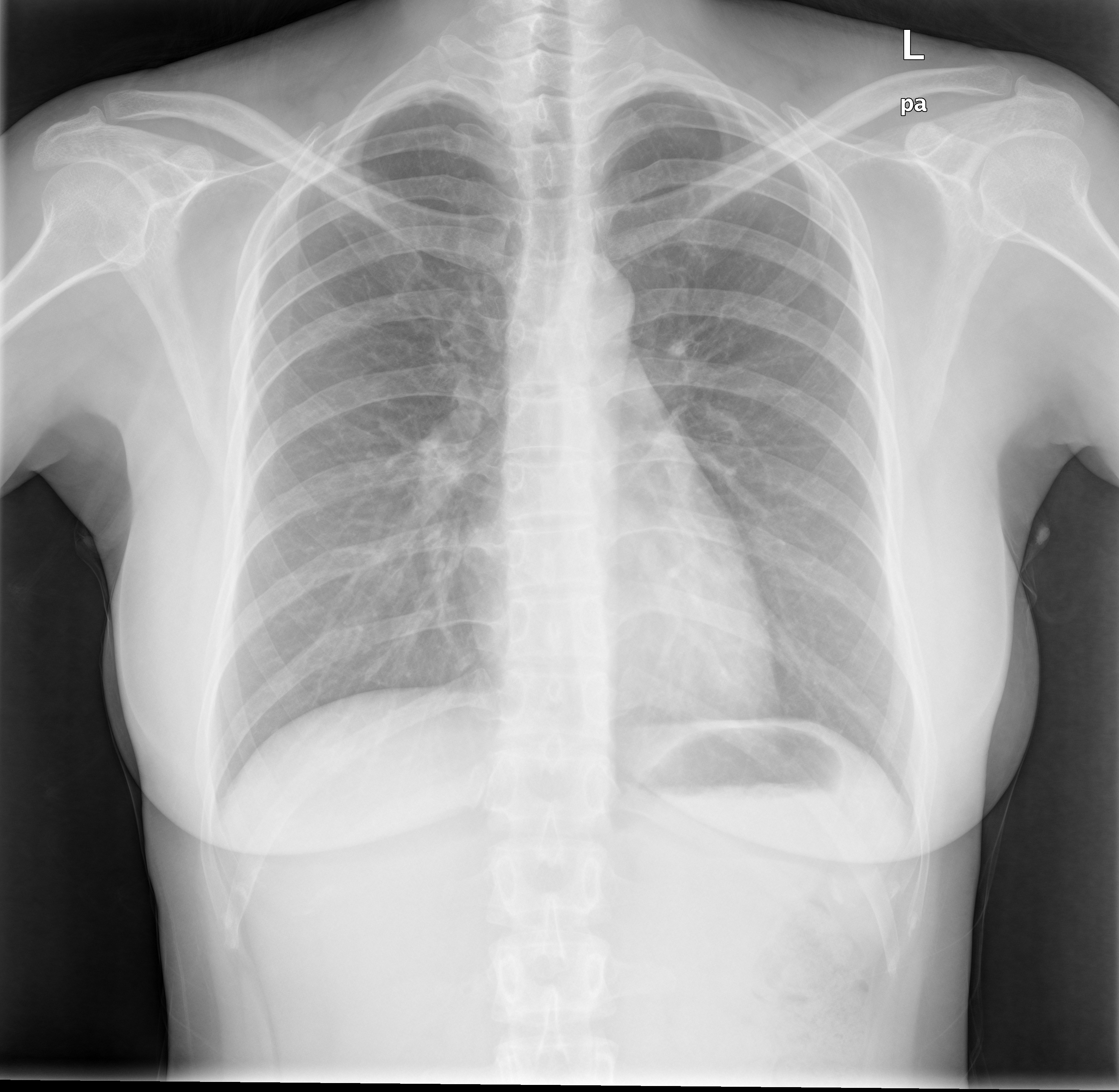} &   \includegraphics[width=30mm]{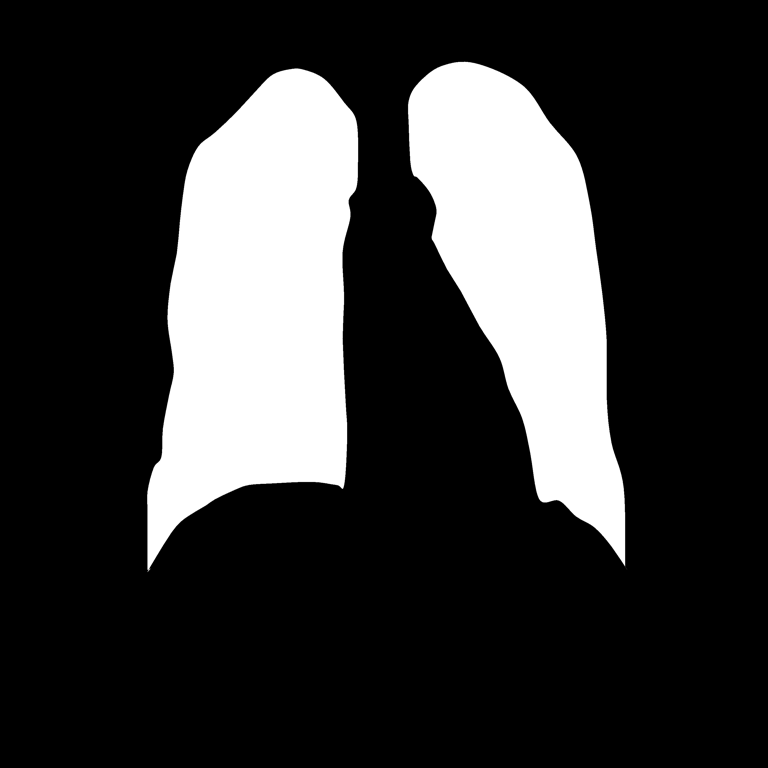}&
  \includegraphics[width=30mm]{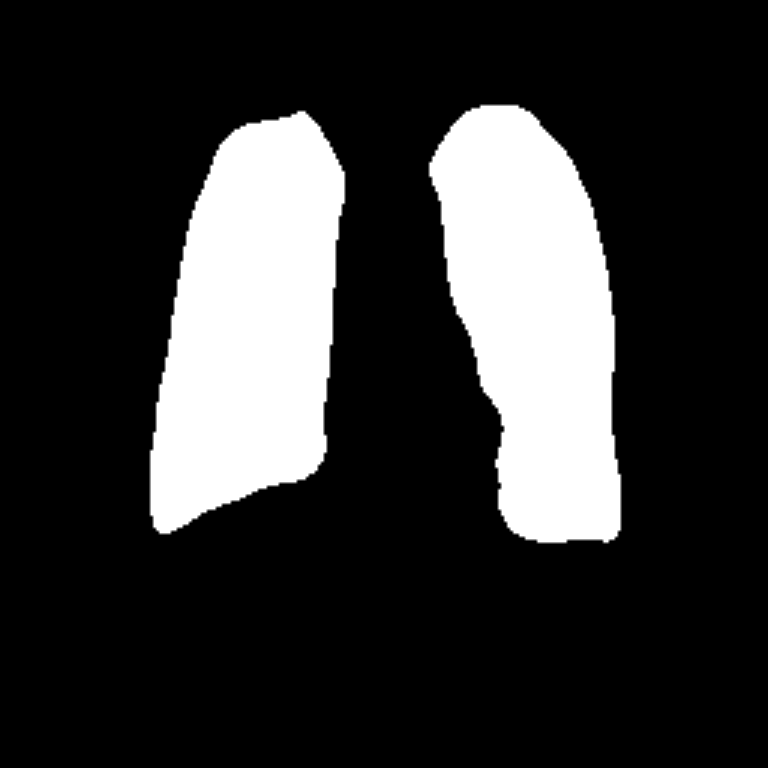}&
  \includegraphics[width=30mm]{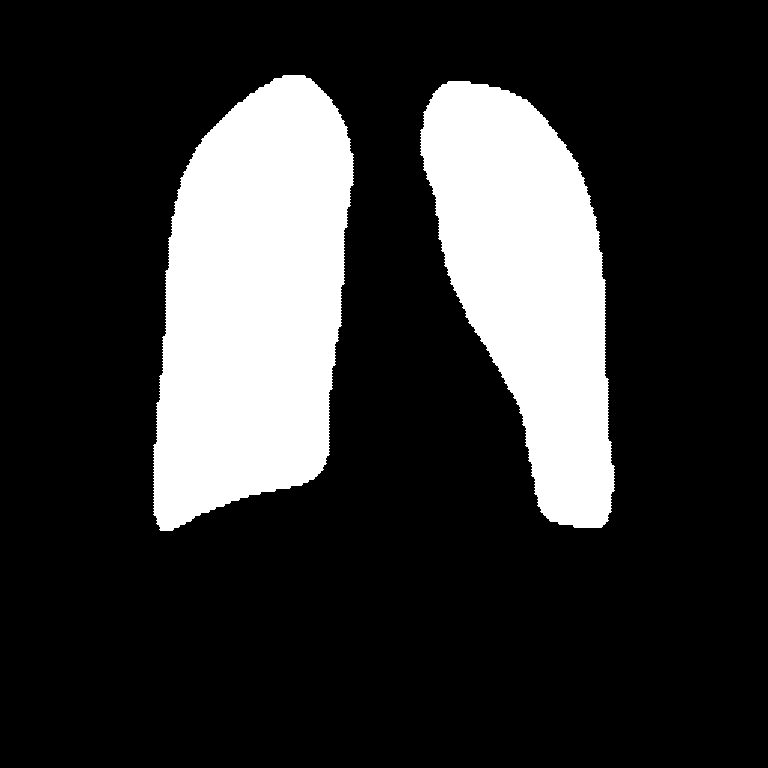}\\
(a) Chest X-ray & (b) ground truth &(c)noise 75 $\%$ baseline  & (d)noise 75 $\%$ ours\\[6pt]
\end{tabular}
\caption{Two examples of the segmentation results in the test data by different methods. (a) is the input image, (b) is the ground truth. (c) and (d) shows the results of U-Net and our method under 75$\%$ noise ratio. }
\end{figure}

\subsection{Quantitative Evaluation}
\begin{table}[t]
\centering
\caption{Comparison between our method and various methods.}
\label{tab:tab1}
\begin{tabular}{l|l|l|l|l|l|l}
\hline
Noise ratio & Noise level & Strategy & Dice & \multicolumn{3}{l}{k} \\ \cline{5-7}
&&&&20&50&80\\ \hline
No noise         & -           & Vanilla U-Net                              & 0.8989 &-&-&-\\
No noise          & -  & Co-teaching \cite{han2018co} &0.9146 &-&-&- \\
No noise         & -           & SS + JO &- &\textbf{0.9252} & 0.9250 &0.9236 \\ \hline
25$\%$           & $5\leq n\leq 15$      & Vanilla U-Net                              & 0.8758 &-&-&-\\
25$\%$          & $5\leq n\leq 15$ & Co-teaching \cite{han2018co} &0.8906 &-&-&-\\
25$\%$           & $5\leq n\leq 15$      & SS                    & 0.9142 &-&-&-\\
25$\%$           & $5\leq n\leq 15$      & SS + JO & -&0.9214 &\textbf{0.9253} &0.9248  \\ \hline
50$\%$           & $5\leq n\leq 15$      & Vanilla U-Net                              & 0.8478 &-&-&-\\
50$\%$          & $5\leq n\leq 15$ & Co-teaching \cite{han2018co} &0.8896 &-&-&- \\
50$\%$           & $5\leq n\leq 15$      & SS                    & 0.8956 &-&-&-\\
50$\%$           & $5\leq n\leq 15$      & SS + JO &- & 0.8790 & \textbf{0.9003} &0.8862 \\ \hline
75$\%$           & $5\leq n\leq 15$      & Vanilla U-Net                              & 0.8496 &-&-&-\\
75$\%$          & $5\leq n\leq 15$ & Co-teaching \cite{han2018co} &0.9023  &-&-&-\\
75$\%$           & $5\leq n\leq 15$      & SS                    & 0.9041 &-&-&-\\
75$\%$           & $5\leq n\leq 15$      &SS + JO &- & 0.9065 & \textbf{0.9108} &0.9067 \\ \hline
\end{tabular}
\end{table}
 The experiments were conducted on the Chest X-ray dataset. We trained the network on the samples with different levels of noisy labels and tested it by the clean labels. Table 1 presents the segmentation performance of vanilla U-Net (baseline) and our cascaded robust learning framework that were all trained by noisy labels. We first trained the fully supervised vanilla U-Net with the noisy level set to zero, which can be regarded as the upper-line performance. Compared with the vanilla U-Net, our framework improves the segmentation performance and achieves an average Dice of $0.925$ on the clean annotated dataset, indicating that the sample selection stage and joint-optimization stage can actually encourage the model to learn more distinguish features. 

For the training dataset with different level of noisy labels, we observed that as the noise level increases, the segmentation performance of the vanilla U-Net decreases dramatically. Compared with vanilla U-Net, the sample selection stage (SS) can consistently improves the performance by encouraging the model to be trained by the selected data. Through the joint optimization (JO) stage supervised by the corrected label and original ones, the segmentation accuracy are further improved, suggesting that our method can effectively eliminates the effect of the noisy and gain performance by producing correct label. In Fig. 2, we show some segmentation results under 75$\%$ noise, in which our results have higher Dice score than the baseline method. At all the noise level, we compared our method with the state-of-the-art noise robust method \cite{han2018co}, which select the small loss samples according to the prediction of peer network. The results show that our method outperform the state-of-the-art method in all the noise level setting. 

In our experiment, we also investigated the impact of the starting epoch $k$ on the performance of our method.
As shown in Table 1, the joint optimization (JO) with label correction stage is started at 20, 50, and 80 epochs, respectively. The experimental results show that the segmentation has the best accuracy at the intermediate epochs.

\subsection{Analysis of Our Method}
\begin{figure}[t]
\label{fig:comp}
\begin{tabular}{ccc}
  \includegraphics[width=40mm]{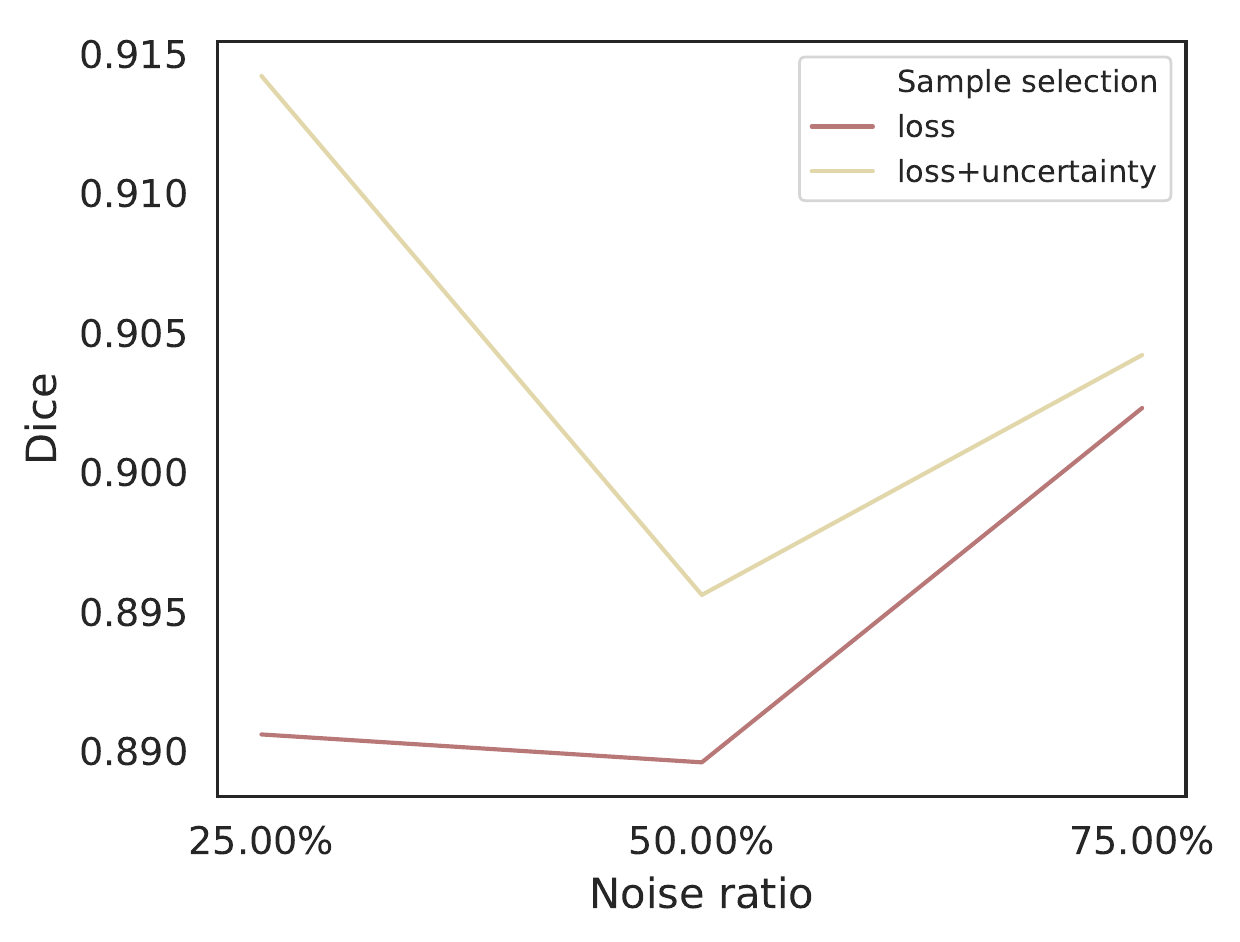} &
  \includegraphics[width=40mm]{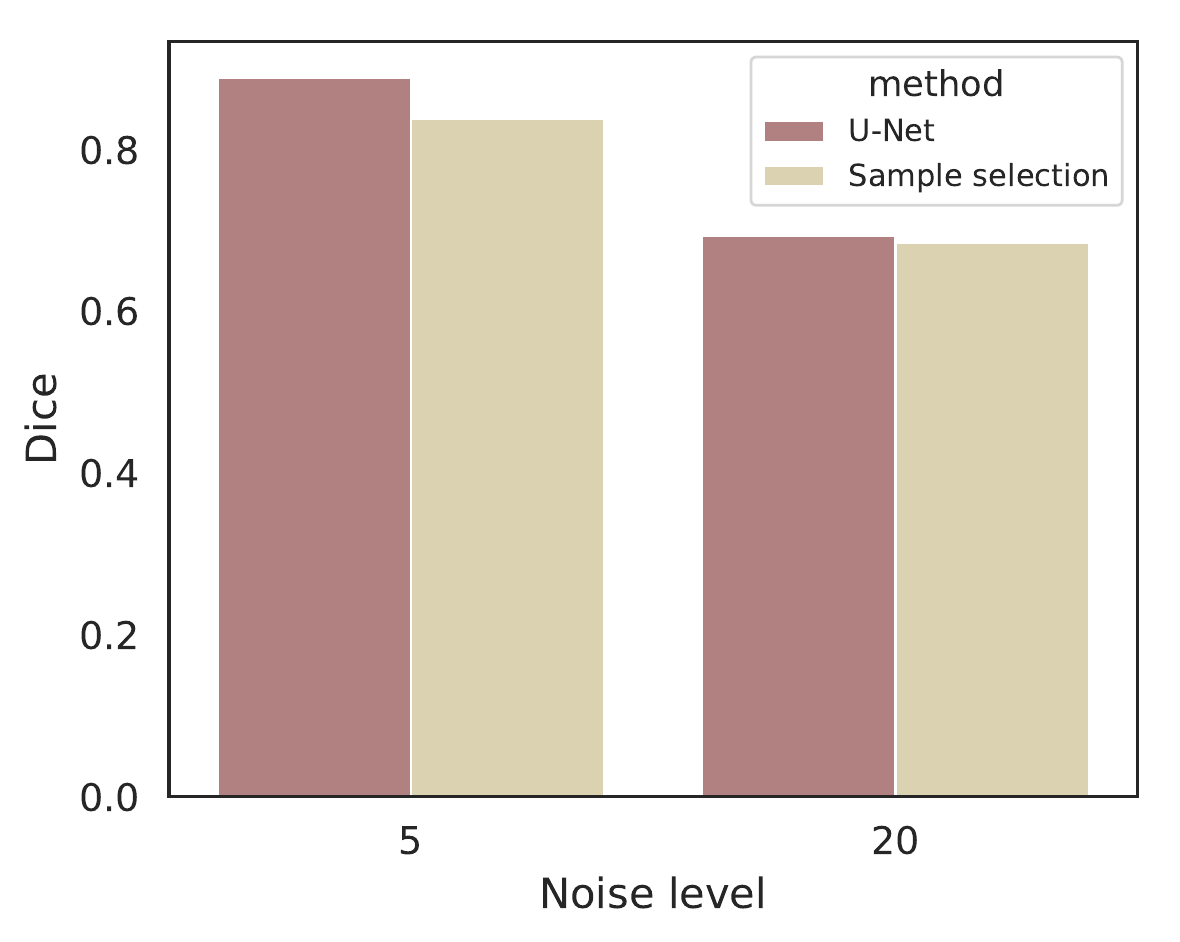} &   \includegraphics[width=40mm]{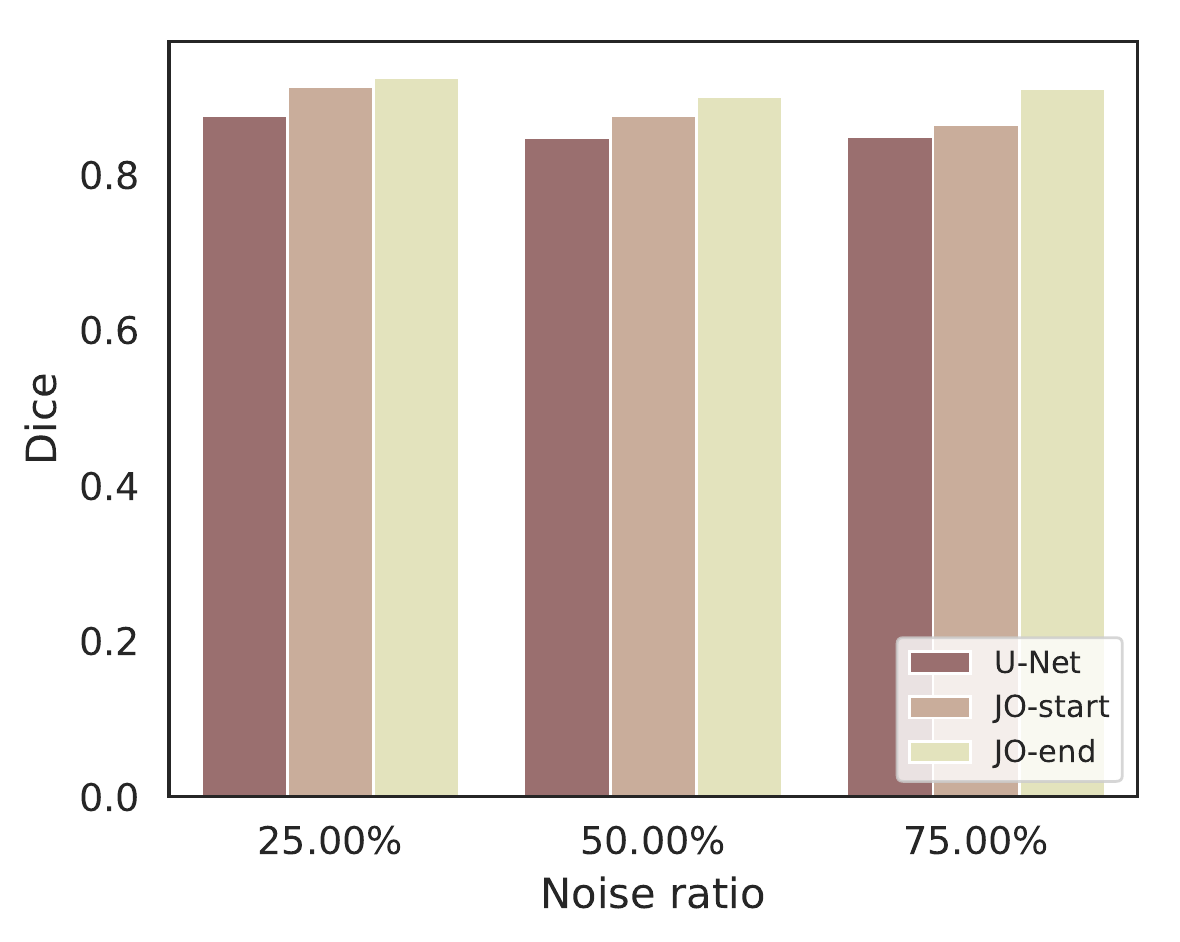}\\
(a)  & (b) & (c) \\[6pt]
\end{tabular}
\caption{(a) The segmentation accuracy of different sample selection criteria. (b) The segmentation accuracy of the U-Net and U-Net with only sample selection stage on 100$\%$ noise setting. (c) The label accuracy of labels in the original dataset , and labels corrected by the model at the end of training.}
\end{figure}


\paragraph{Sample selection}
Compared with the vanilla U-Net, our sample selection stage (SS) shows higher segmentation accuracy under different noisy level, as shown in Table 1. To validate the criteria of our sample selection, we conducted another experiment that only selected the small loss sample. Fig.3(a) shows the test accuracy with different sample selection criteria. It reveals that the test accuracy significantly improved when considering the uncertainty in the selection stage. To further validate the effectiveness of our method at the sample selection stage, we applied our method on training dataset with 100$\%$ noise and noise level $n=5,20$. Under this setting, the sample selection stage shows worse segmentation accuracy than vanilla U-Net, because no clean sample can be selected. The results decreased due to the low sample utilization efficiency.  

\paragraph{Joint optimization}
To analyze the contribution of the joint optimization stage, we explore the label accuracy with and without the stage of joint optimization and label correction. We calculated the Dice coefficient of the initial noisy label ($\hat{y}$) and the corrected label ($\bar{y}$) of the final model at the end of training. Fig.3(c) shows the overall accuracy. We see that the label quality is improved by the scheme of joint optimization and label correction. 

We also investigated the impact of the weight factor $\alpha$ by adjusting $\alpha$ between 0 and 1. If $\alpha=0$, the network is trained only using the noisy labels without correction, and $\alpha=1$ represents the network discards the original label and only use the corrected labels. The test accuracy with different $\alpha$ under 50$\%$ noise ratio is shown in Fig 3.(d). Results show that model trained jointly by the corrected label $\bar{y}$ and original label $\hat{y}$ achieves the best performance when $\alpha=0.5$. $\alpha=1$ leads to sub-optimal performance as it might correct some hard samples, and thus reduced the network generalization ability.

\section{Conclusion}
In this paper, we present a novel Cascaded Robust Learning framework for the segmentation of noisy labeled chest x-ray images. Our method consists of two stages: sample selection stage, and the stage of joint optimization with label correction. In the first stage, the clean annotated samples are selected for network updating, so that the influence of noisy sample can be interactively eliminated in the three networks. In the second stage, the label correction module work together with the joint optimization scheme to revise the imperfect labels. Thus the training of whole network is supervised by the corrected labels and the original ones. Compared with other state-of-the-art models, our cascaded robust learning framework keeps high robustness when the training data contains various noisy labels. Experimental results on the benchmark dataset demonstrate that our network outperforms other methods on segmentation tasks and achieves very competitive results on the noisy-labels dataset.

%
%
%
\bibliographystyle{splncs04}
\bibliography{ref}

\end{document}